# Influence of Chain Structure and Swelling on the Elasticity of Rubbery Materials: Localization Model Description

Jack F. Douglas

Polymers Division, National Institute of Standards and Technology, 100 Bureau Drive, Gaithersburg, MD 20899-8542, USA
Fax: (301) 975- 4924; E-mail: jack.douglas@nist.gov

**Summary:** Classical network elasticity theories are based on the concept of flexible volumeless network chains fixed into a network in which there are no excluded volume, or even topological, interactions between the chains and where the chains explore accessible configurations by Brownian motion. In this type of 'classical' model of rubber elasticity, the elasticity of the deformed network derives from the entropic changes arising from a deformation of the network junction positions. The shortcoming of this approach is clear from the observation that unswollen rubbery materials are nearly incompressible, reflecting the existence of strong intermolecular interactions that restrict the polymer chains to an exploration of their local tube-like molecular environments. The imposition of a deformation of these solid rubbery materials then necessitates a consideration of how the local molecular packing constraints become modified under deformation and the impact of these changes on the macroscopic elasticity of the material as a whole. Many researchers have struggled with this difficult problem, in the present paper we focus on the simple 'localization model' of rubber elasticity of Gaylord and Douglas (GD), which provides a simple minimal model for the network elasticity of rubbers having strong intermolecular interactions in the dense polymer state. Particular emphasis is given in the implications of this model in describing how network elasticity changes with swelling by a solvent, a phenomenon where large deviations from classical elasticity have been observed and a situation relevant to numerous applications involving rubbery materials. We also discuss the nature of entanglement based on the same packing picture and deduce general relationships for entanglement in terms of molecular parameters and we find that our predictions accord with recent experimental correlations relating chain molecular structure to the entanglement molecular mass.

The essential idea of the localization model of rubber elasticity is that that the network chain segments are localized at the network junctions, as in classical elasticity theory, but they also consider the chain segments to be localized along their contours by a inter-particle packing interactions, as in atoms in a crystal. This picture of the inter-chain interaction is the familiar Edwards-De Gennes 'tube' model of polymers in the condensed state. The existence of the tube means that the entropy of the network chains is reduced relative to chains without this constraint, but the crucial problem is how this entropy change becomes *modified* as the material becomes deformed. GD approach this problem by first observing that both the volume of the material and the network chains are essentially invariant under deformation so they require that the dimensions of the confining tube to also be a deformation invariant where this constraint is applied at a segmental level.

## Review of the Localization Model of Rubber Elasticity

Models of network elasticity have concentrated on minimal aspects of rubber networks. In particular, the classical rubber elasticity models of Wall and Flory, [1] James and Guth, [2] Edwards [3] and many others have focused on the consequences of network connectivity. Classical theories idealize rubbery materials as idealized networks of random walk chains whose junctions on average deform affinely in response to a macroscopic deformation, thereby changing the entropy of the system. More recent work has emphasized inter-chain interactions or 'entanglement' interactions defined in terms of the topological constraint of chain un-crossability and correlations arising from molecular packing of the network chains, A minimal statistical mechanical model of rubber elasticity must incorporate three main features of the network chains: 1) A connected network of flexible chains. [1-3] 2) 'Entanglement' constraints. [4-7] 3) Finite volume of chains. [8] The localization model (LM) of rubber elasticity directly addresses these effects. [9,10]

As a first approximation, the localization model (LM) takes the free energy $\Delta F_{network}$ of the network per unit volume to be proportional to the number of chains per unit volume $\nu_d$ where $\Delta F_{chain}$ is the chain free-energy of deformation, $\Delta F_{network} \sim \nu_d \Delta F_{chain}$. This relationship, which can be derived from classical rubber elasticity theory, assumes that each chain sees an *equivalent* molecular environment arising from its interaction with surrounding chains and this approximation thus amounts to a mean field approximation [11] when it is applied more broadly. For a cross-linked network, the number of chains per unit volume is taken to be proportional to the number of cross-links per unit volume, the cross-link density, $\nu_d$. The second basic approximation of LM is to assume that the network chains are Gaussian chains in our estimation of the chain connectivity contribution to $\Delta F_{network}$. This approximation has its limitations, especially for large network extensions and for cases where the network chains are stiff.

Before modeling of the entanglement contribution to $\Delta F_{network}$, we make some general physical observations that constrain our theoretical development:

1) Dry rubbery materials are normally nearly incompressible because of strong repulsive inter-segment interactions, despite the random coil nature of the polymers.

2) Confinement of network chains to a volume on the order of the hard core volume of the chains alters the average chain entropy relative to an unconfined chain.

The question is then how you calculate this entropy change under deformation conditions?

The Feynman-Kac functional limit theorem (FKLT) implies that confining a chain by *any* means leads to a *universal* change in the chain free energy of the flexible polymer chains due to confinement, $F_{conf} \sim <R^2> / \xi^2$, where $<R^2>$ is the mean squared dimensions of the unconfined chain and $\xi$ is the *localization length* describing the scale over which the chain is localized.[12] $<R^2>$ is proportional to the chain length, N, so that $F_{conf}$ is extensive in the chain length. This limiting scaling relation is also known as 'ground state dominance' [13] based on a quantum analogy with the Brownian chain model. More generally, we have the more general scaling relation, $F_{conf} \sim <|R|>^{df} / \xi^{df}$ for generalized random walks (having independent steps, but whose variance in step length is not finite) where $<|R|>$ is the mean chain size and $d_f$ is the fractal dimension of the chain, i.e., $<|R|> \sim N^{1/df}$. Evidently, we again obtain a confinement contribution $F_{conf}$ that is extensive in N. In the presence instance, this localization effect derives from the hard core repulsive interactions between a given chain and its surrounding chains so that $F_{conf}$ is entropic in nature. In other words, hard core excluded volume interactions confine the chains to 'tube-like' regions localized around some average chain conformation and this chain confinement gives rise to a change of the entropy per link of the polymer chain. The FKLT provides the fundamental mathematical underpinning of the tube model of polymer melts and rubber elasticity.

So far we have based our model on two fundamental limit theorems of broad mathematical and physical significance- the central limit theorem describing the statistical properties of random walk chains and the FKLT describing how the free energy of these chains changes with confinement (or the mathematical equivalent of this effect). This provides a sound foundation for a general theory of flexible polymer networks with strong localizing inter-chain interactions in the melt state. The problem of calculating how $F_{conf}$ becomes modified by macroscopic deformation of the rubbery material is more difficult. It seems reasonable to assume that under quasi-equilibrium conditions of deformation that the FKLT relation still applies and the crux of the LM modeling then reduces to estimating how $\xi$ varies with deformation. There is certainly no reason to believe that $\xi$ should vary affinely with the extent of macroscopic deformation $\lambda_i$ along the laboratory-fixed axes, as might reasonably be argued for the coordinates defining the chain junctions in the network.

Gaylord and Douglas [9] approached this basic problem by assuming, as Edwards[14] had done before, that the network chains are contained within a random tube with local harmonic confining potential that is composed of segments that are oriented along three directions (x, y, z) in the lab-fixed frame. The harmonic tube model for the inter-chain interaction potential is chosen simply for mathematical convenience. The FKLT theorem ensures that essentially any reasonable confining potential will lead to the same limiting results. We note that Heinrich et al. [15] have also developed a popular tube model of rubber elasticity, based on the same chain localization concept, but these authors do not invoke the packing arguments to specify the molecular parameters in their model and change in elasticity with network swelling.

The random tube model can now be constructed by viewing the random tube as consisting of straight tube sections lying along the macroscopic deformation axes. The distribution function describing the distribution of the chain monomers within a random tube segment then factorizes into a product of Gaussian functions defined in terms of coordinates along the tube axis and a coordinate normal to the tube axis. Because of the separability property of random chains the random tube segments can be imagined to be aligned along the three macroscopic deformation axes with equal probability, a construction first introduced by James and Guth in their approximate treatment of finite-extensibility effects on network elasticity.

To calculate the free energy change with deformation, the junction positions are taken to deform affinely, $R_i = \lambda_i \cdot R_{io}$, where $R_{io}$ is the initial distance between a given network chain end along the $i^{th}$ macroscopic deformation coordinate direction. This argument leads to the classical affine network model of rubber elasticity. Of course, the affine deformation assumption is an approximation and other models of rubber elasticity take this as a starting point of their development.[16] It is this author's opinion that non-affine contributions do not really address the inter-chain interaction effects responsible for 'entanglement' contributions to the elasticity of dry rubbers.

To estimate $\xi(\lambda_i)$, we argue that the hard core volume of the chain and localizing tube are comparable and invariant to deformation. The assumption of affine displacement is taken to mean that that length of the tube segments along the deformation axis deform affinely,

$$L = \lambda_x L_o \, ,$$

where $L_o$ is the length of the undeformed tube. The invariance of the tube volume with this deformation implies that the length of the tube times its cross-sectional area is the same before and after deformation, $L \, \xi_x^2 = L_o \, \xi_{xo}^2$. This relation implies that $\xi_x = \lambda_x^\beta \, \xi_o$, where $\beta = -1/2$ and $\xi_{xo} = \xi_{yo} = \xi_{zo} \equiv \xi_o$. This scaling relation is obviously quite different from an affine variation, $\xi_x = \lambda_x \, \xi_o$. Rubinstein-Panyukov [17] have made arguments that $\beta$ is positive ($\beta = \frac{1}{2}$), which is a similar scaling found based on assuming that scales also scales affinely so that $\beta = 1$. The model of Rubinstein-Panyukov is reasonably based on the assumption of topological invariance of the network under macroscopic deformation, but models the polymers as volumeless filaments so that this model does not account for chain packing effects. Note that the LM and the Rubinstein-Panyukov models differ in the sign of $\beta$ so we are not talking about a subtle difference in the model predicttions.

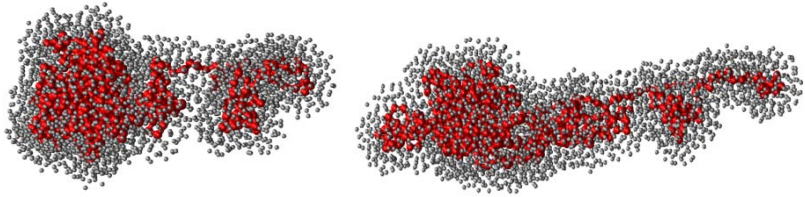

Figure 1. Molecular dynamics simulation of chain deformation under macroscopic extension (the macroscopic deformation direction (x) is along the horizontal direction; left chain is before deformation and the right is after deformation. Red spheres indicate chain segments and the gray segments are those of surrounding chains in the proximity of the illustrated test chain. The number of chain segments in the test chain environment is nearly invariant- consistent with an invariance of the tube volume with deformation. Figure is for schematic purposes only.

These considerations above then lead to the LM model [9] expression for the free-energy density of a dry rubber,

$$\Delta F_{LM} = (G_c \, / \, 2) \, \Sigma_{x,y,z} \, (\lambda_i^2 - 1) + G_e \, \Sigma_{x,y,z} \, (\lambda_i - 1) \qquad (1a)$$

where the classical network theory shear modulus $G_c$ is proportional to the cross-link density ($\nu_x$) and thermal energy ($k_B T$),

$$G_c = C_o \, \nu_x \, k_B T, \qquad (1b)$$

In the affine network model of Wall and Flory the prefactor is $C_o = 1$ and in the classical non-affine 'phantom model' of James and Guth, [2] Deam and Edwards, [3] where the network junctions can fluctuate with the constraint of excluded volume interactions, $C_o = \frac{1}{2}$. More generally, $C_o$ depends on the details on network structure (dangling ends, network functionality, etc.), [18] and $C_o$ is considered to be a measurable parameter characterizing a given network. [19] In the absence of other information, we take the 'phantom' elasticity model estimate, $C_o \approx \frac{1}{2}$, as an estimate of the 'front factor'. The entanglement contribution $G_e$ to the free energy density of the network,

$$G_e = \gamma \ G_c + G_N^{\ *}, \tag{1c}$$

which includes a cross-term $\gamma$ proportional to $G_c$ [and thus $\nu_x$; see Eq. (1b)] and a cross-link independent term that is identified with the plateau modulus of the polymer melt, $G_N^{\ *}$.[9] Much is known about the dependence of $G_N^{\ *}$ on molecular parameters.[20] We note that Eq. (1a) is consistent with the Valanis-Landel separable form of the strain energy density of rubbery materials, a property of the strain energy density that has been established to be a good approximation for many rubbery materials.[21] The Valanis-Landel property greatly simplifies calculation of the deformation properties of rubbers.

This expression implies the remarkable result that for polymers with a low cross-link density (so that the rubber is barely a solid) elastic modulus reduces to that of a polymer melt and that the cross-links have rather little contribution to the elasticity. At high cross-link densities, the shear modulus of the rubber scales in linear proportion to $\nu_x$ as in classical elasticity theory.

## Application of the Localization Model to the Elasticity of Dry Rubbers

Calculations of the stress - strain relations for dry rubber are then remarkably simple for the LM model.[9,22] Under uniaxial deformation($\lambda_x = \lambda$ ; $\lambda_y = \lambda^{-1/2}$; $\lambda_z = \lambda^{-1/2}$ ) of an incompressible material in three dimensions, we have the stress $\tau$,

$$\tau \ = \ d \ [\Delta F_{network} \ / \ V_o] \ / \ d \ \lambda \ , \tag{2}$$

where $V_o$ is the dry rubber volume. Eq. (2) then implies the simple relation,

$$\tau \ = G_c \ (\lambda - \lambda^{-2}) + G_e \ (1 - \lambda^{-3/2}). \tag{3}$$

The stress relative to its classical variation ($\lambda - \lambda^{-2}$) defines a reduced stress:

$$I \ (\lambda) \equiv \tau \ / \ (\lambda - \lambda^{-2}) = G_c + G_e \ (1 - \lambda^{-3/2}) \ / \ [ \ \lambda \ (1 - \lambda^{-3})]. \tag{4}$$

For large deformations ($\lambda \rightarrow \infty$) this expression reduces to an asymptotic *Mooney-Rivlin* relation,[23]

$$I \ (\lambda) \sim G_c + G_e \ (1 \ / \ \lambda) \ . \tag{5}$$

so that $G_c$ and $G_e$ can be identified approximately with the Mooney parameters $2 \ C_1$ and $2 \ C_2$ that are normally considered in characterizing dry rubbers under extension. Blokland[23] has provided an extensive tabulation of $C_2$ for a variety of rubbers at relatively high cross-link densities and we tentatively suggest the approximation $\gamma \approx 1/3$ as a useful approximation in the absence of direct measurement.

The reduced stress in the LM model can be generalized to d-dimensions:

$$\tau = G_c \ [\lambda - \lambda^{\ -(d+1)/(d-1)}] + [2 \ G_e \ / \ (d-1)] \ [ \ \lambda^{2/(d-1) \ - \ 1} - \lambda^{\ -2/(d-1)(d-1) \ - \ 1} ] \ , \tag{6}$$

and this quantity reduces to the classical–like expression $\tau = G_c \left[ \lambda - 1/\lambda \right]$ in high dimensions [25] to an expression applicable to the elasticity of rubbery membranes (e.g., polymerized Langmuir films) for d = 2.[26]

Previous comparisons of the LM model to experiment show that this simple model compares very well with uniaxial compression, extension measurements, as well as biaxial extension, [9] and torsional deformation measurements on dry rubbers. [27] Moreover, the LM also performed rather well on the complex problem of describing the elasticity of rubbers that have been cross-linked in a deformed state. [28] The LM is critically compared to other models by Han et al. [29] and illustrate fits of the localization model to well known careful observations on dry rubber elasticity are given by Lin et al. [30], where the theory is also extended to describe stiff fiber networks and networks formed by self-assembly.

The torsional deformation measurements of McKenna et al. [27] provide a particularly notable test of the LM since they were performed over a wide range of cross-linking density. An analysis of these measurements (where $C_o$ is fixed to ½) yielded a $G_e$ having exactly the form predicted by Eq. (1), i.e., $G_e$ is linear in $v_x$ and extrapolates to $G_N^*$ in the limit of vanishing cross-linking density. These observations are highly reassuring regarding the physical basis of the LM model. We next consider the extension of the LM model to describe the change in elasticity of rubber subjected to swelling.

## Localization Model Description of the Elasticity of Swollen Rubbers

The theoretical prediction of the elasticity of swollen networks from information about the dry rubber is a challenging problem. Before addressing this problem, we note some insightful comments and observations made by Gumbrell et al. [8] on this topic:

> "The change in $C_2$ with volume swelling can be associated with the finite volume of the rubber molecules. This leads to a reduction in the number of possible configurations as two molecules cannot occupy the same space at the same time nor can they pass through one another. The reduction of configurations from this cause would naturally be less in the swollen than in the dry state and in highly swollen rubbers deviations from ideal behavior due to this cause will be small."

> "The value of $C_2$ is found to be large in dry rubbers and decreases to zero at high degrees of swelling."

We must first address, in a way consistent in a way consistent with the formulation of the LM model, the concentration dependence of the plateau modulus $G_N^*$ of a polymer melt. In a melt of high molecular mass polymers, the chains are only *transiently* localized into their local tube environments on the time scale of the stress relaxation time, the terminal time. The system thus responds elastically when perturbed at relatively high frequency measurements, while the system flows over long time scales. Cross-linking locks chains into a *permanently localized state*- an amorphous solidification transition. [31]

In the limit $\nu_d \rightarrow 0$ (i.e., a high molecular mass polymer melt), the entanglement contribution $G_e$ is simply due to chain localization due to inter-chain interactions:

$G_e(\nu_d \rightarrow 0) \approx G_N^*$ [9,12] where $\Delta F_{local} \sim G_N^* \sim \nu_d <R^2> / \xi^2$ . If we imagine the network as being comprised of one single molecule (tube) of length N that fills space, then the change in system volume upon swelling $V_o \rightarrow V$ implies the change in the localization length, $\xi \sim \xi_o \phi^{-1/(d-1)}$, where $\phi$ is the polymer volume fraction and the cross-link density as, $\nu_d = \phi \nu_{do}$. These relations, which derive directly from the consideration of space filling of the network chains, then imply that the plateau modulus of the diluted melt $G_N(\phi)$ scales as,

$$G_N(\phi) = G_N^* \phi^{(d+1)/(d-1)} \text{ or } G_N(\phi; d=3) \sim G_N^* \phi^2 . \qquad (7)$$

It is emphasized this scaling has nothing to do with the fractal character of the chains.

## Molecular Parameters Governing $G_N^*$ and the Chain 'Entanglement'

These simple packing arguments have strong implications for the dependence of network chain monomer structure and the extent of network cross-linking- effects not at all considered in the classical network elasticity models. In particular, this same physical picture implies that $G_N^*$ should scale in rough inverse proportionality to the chain cross-sectional area $A$ [32,33] since the network chains in the dry rubber are confined to a scale comparable to the segment size so that the localization length $\xi$ should scale as $\xi \sim A^{1/2}$. [12] (We return to this point below where establish this relation quantitatively.) This simple packing argument implies that the entanglement contribution to the elasticity of polymer networks, and even polymer melts, should be reduced in chains with bulky side-groups that increase the effective chain cross-sectional area , $A$. To achieve such an effect, however, it is also necessary for the chain spatial extent of the chains to be large in comparison to their chain cross-sectional radius so that they are coupled in their motions. This reminds us that polymers generally exhibit a *transition* between an 'unentangled' and 'entangled' state with increasing chain length, raising questions about the nature of entanglement and about what molecular parameters govern this transition. This problem requires a consideration of physical nature of the polymer packing interactions in the condensed state that were neglected in the original formulation of the LM. It is certainly not obvious that entanglement involves chain knotting, as the term naively suggests, since stiff polymer chains with a limited capacity for knotting tend to exhibit even stronger entanglement effects than flexible polymer chains. [34-37] The terms 'entanglement' and gel are applied loosely to suspensions of such rigid and extended objects [38] as multi-wall carbon nanotubes and suspensions of other fibers [39] and even exfoliated clay suspensions. [40] We seek a definition of entanglement that can address these systems within a unified and predictive framework. It is evident that the physically appealing mean field picture of 'reptation', in which chains are assumed to 'crawl' out of the tube defined through its interaction with surrounding chains, is inadequate for modeling the many-body interactions arising within disordered configurations of fibers and sheets that give rise to emergent rheological properties that that can be quite different from the entangled particle systems when the particle concentration is sufficiently high and particle anisotropy is sufficiently high. 'Entanglement' seems to involve the onset of some sort of collective particle motion and a consideration of this problem requires a more general consideration of the physical conditions under which emergent collective motion arises in extended objects exhibiting large and anisotropic excluded volume interactions.

Douglas and Hubbard [12, 41] developed a model of 'entanglement' based on the conception that inter-chain correlations associated with packing and topological interactions between the chains leads to groups of chains moving collectively in the form of polydisperse chain *clumps* that form and disintegrate in dynamic equilibrium. Specifically, Douglas and Hubbard [12] argue that etanglement under equibrium conditions corresponds to a type of supermolecular assembly transition that is entropy driven [42], as in the crystalliztion of hard spheres [43,44] and the liquid crystaline ordering of rod and sheet-like molecules having a sufficiently large aspect ratios. [45-48] This view of entanglement should apply not only to flexible and semi-flexible linear polymer chains, but to many other types of extended particles having sufficient configurational disorder and/or size polydispersity that liquid crystalline ordering cannot arise. Self-assembly of molecular has similarly been suggested to underlie glass-formation where the presence of configurational disorder and other disordering effects such as impurities make crystallization no longer possible so that self-assembly, a weaker sort of ordering, occurs by default. [12, 49] It is then reasonable from this perspective to speak of entanglement in carbon nanotubes and fibers, as well as in exfoliated clay suspensions where strong anisotropic excluded volume interactions still prevalent so that particle motion must be collective, involving the movement of many particles at once. The dynamic particle clustering implied by this reasoning has recently been observed in measurements of driven granular rod systems and simulations where spherical particles were introduced to create sufficient disordering effect to suppress nematic ordering and where the dynamic clusters were reasonably described as a kind of equilibrium polymerization. [50] Notably, an ordinary isotropic-nematic transition with increasing rod concentration was observed without the added spheres.

It is frequently observed that nematic ordering of rod-like molecules requires a critical ratio of the rod length to diameter ratio of about $x_c \approx 5 \pm 1$ (e.g., see refs. [51] and [52]). This critical aspect ratio for long range ordering is reasonably close to the theoretical estimates of this critical aspect ratio (Flory's original lattice model estimate of 6.7 [46], but later more refined estimates of this ratio were derived by DiMarzio [47] and Freed and coworkers [48] indicated a lower value near 4.) Douglas and Hubbard [12] reasoned, by extension of these well-known results for rod ordering, that the same geometrical inter-polymer interaction condition should describe 'entanglement' where the rod length is suitably replaced by a length on the order of the chain span or more simply the root mean square chain dimensions $<R^2>^{1/2}$ (Even flexible polymer chains exhibit significant anisotropy in their average shape and this anisotropy is even greater in stiffer chains). This simple proposition provides precise and testable predictions for the molecular characteristics of polymer chains that lead to chain entanglement that are applicable to both flexible and semi-flexible polymers.

Taking $<R^2>^{1/2}$ to be the on the order of the extended dimension of the rod and the condition $x_c \approx 5$ gives then leads to the chain 'entanglement' condition [12]

$$x_c^2 \ = \ <R^2>_c \ / \ D^2 \ = (<R^2>_c \ / \ M_c) \ M_c \ / \ D^2 \ = \ 25 \tag{9}$$

so that the critical molecular mass $M_c$ becomes,

$$M_c = (32) \ A \ / \ (<R^2> \ / \ M) \ \sim \ A, \tag{10}$$

where we assume a circular chain cross-section so that $A = (\pi \ /4) \ D^2$ and where $(<R^2>_c \ / \ M_c)$ can be reasonably approximated by the widely tabulated solution property $(<R^2>_\theta \ / \ M)$. He and Porter [33] experimentally examined the critical conditions for entanglement for about 40 polymers and independently arrived at Eq. (10), except the constant 32 in this equation was found instead to be closer to 28 or,

$$M_c \approx (28) \ A \ / \ (<R^2>_\theta \ / \ M). \tag{11}$$

A similar geometrical correlation between entanglement and the geometrical properties of the individual chains was found by Zang and Carreau. [53] There is evidently strong experimental evidence in support of the physical picture of necessary conditions for chain entanglement introduced by Douglas and Hubbard. [12]

The critical conditions for chain entanglement transition, which is phenomenologically signaled by a much larger mass scaling exponent relating the shear viscosity and structural relaxation time to the polymer mass and to the onset to strong shear thinning under steady shear [54-56] in terms of molecular parameters can be refined by considering experimentally accessible molecular parameters relating to the space filling nature of polymer chains. Specifically, if we start from the condition for entanglement given by Eq. (11) and note that the chain volume $v_c$ can be estimated from the ratio $v_c \equiv M / \rho$ where $\rho$ is the density and M the chain mass, we obtain a simple scaling relation for the critical molecular mass for chain entanglement, $M_c \sim \rho A \ (v_c / <R^2>)$. Since excluded volume interactions are screened in the polymer melt, the volume of the chain can be described at a coarse-grained level as a sweeping out process of chain segmental units sequentially along the polymer chain. The volume swept out by such a random walk relative to its mean square dimensions $<R^2>$ is a mathematically well-known *functional* of the monomer unit shape called the *'capacity'* $(C)$ [57] and in the polymer literature this ratio between the chain volume and $<R^2>$ has been termed the 'packing length' $p$, which is conveniently a widely tabulated molecular characteristic of polymers describing the space-filling character of the polymer (hence the term 'packing parameter' for $p$). Further, there is a highly a useful and well-supported approximation relating the monomer area $A_{mon}$ to $C$ called the Russell approximation [58], $C \approx (A_{mon} / 4\pi)^{1/2}$, which is applicable provided the monomer is not too irregular in its shape. Thus, we have the alternative scaling relation for $M_c$ that follows from the definition $C = p$ and the scaling relations, $A_{mon} \sim C^2 \sim p^2$,

$$M_c \sim \rho \, p \, A \sim \rho \, p^3 \qquad (12)$$

This accords quite well with the impressive correlations $M_c$ and $p$ indicated in recent work by Fetters and coworkers [59]. This precise description of the geometrical parameters governing entanglement derives naturally from the entropic-self assembly picture described above and defines in a precise sense what it means for the chains to be sufficiently 'long' relative to their cross-sectional dimensions in order for chain entanglement to be exhibited.

We see from Eqs. (1c) and (5) that under moderate extensional rubber deformation, the localization model predicts that the conventional Mooney parameter $C_2$ defined under this particular type of deformation should scale as $G_e$ and for light cross-linking this quantity scales the plateau modulus of the uncross-linked polymer melt $G_N*$. A rough inverse scaling between $C_2$ and the chain cross-sectional area $A$ estimated from scattering measurements has been observed in synthetic rubbery materials. [32, 60] There is clear evidence that the non-classical contribution to the chain rubber elasticity associated with inter-chain interactions is diminished if the network chains are made 'fatter', in basic accord with the packing arguments of the LM. We next consider the specific predictions of the localization model on the elasticity of rubbery materials swollen from the dry state where we focus on limiting conditions where the chain cross-links and the chain interactions have a predominant effect on how the elasticity changes with swelling.

To calculate the tension of a swollen rubber subjected to deformation,[61] we define a swelling factor to describe the first swelling part of the deformation ($V_o \lambda_s^3 = V$), where $\lambda_s$ is swelling factor. This deformation is followed, for example, by a *uniaxial* deformation ($\lambda_x = \alpha$; $\lambda_y = \alpha^{-1/2}$; $\lambda_z = \alpha^{-1/2}$ ) so that the tension $\tau(\alpha)$ is equal, $\tau = d\ [\Delta F_{network}\ /V]\ /\ d\alpha$ or explicitly the LM predicts,

$$\tau = G_c\ \phi^{1/3}\ (\alpha - 1\ /\ \alpha^2) + (\gamma\ G_c + G_N*\ \phi)\ \phi^{2/3}\ (1 - \alpha^{-3/2}) \qquad (13)$$

The concentration dependent reduced stress $I(\alpha,\phi) \equiv \tau\ /\ (\alpha - 1\ /\ \alpha^2)\ \phi^{1/3}$ is then equals,

$$I(\alpha,\phi) =\ G_c + (\gamma\ G_c + G_N*\phi)\ \phi^{1/3}\ (1 - \alpha^{-3/2})\ /\ \alpha\ (1 - \alpha^{-3}) \qquad (14)$$

And in the large extension limit, Eq. (14) formally reduces, as before in the dry rubber case, to the Mooney-Rivlin form,

$$I(\alpha,\phi) =\ G_c + 2C_2(\phi)\ /\ \alpha\ ,\ \alpha \to \infty \qquad (15)$$

where $G_c \equiv 2\ C_1$ and $G_e(\phi) \approx 2\ C_2(\phi)$ and the 'Mooney parameter' $C_2$ exhibits the non-trivial concentration scaling,

$$2C_2(\phi) \approx (\gamma\ G_c + G_N*\phi)\ \phi^{1/3}. \qquad (16)$$

Although $C_2$ generally vanishes upon swelling, as noted by Gumbrell et al. [8] , we see that there can be a *qualitative* difference in the concentration dependence of the elasticity of rubbery materials that have relatively high and low cross-linking densities where first and second terms on the right hand side of Eq. (11) dominate, respectively. For lightly cross-linked materials where the chain localization term related to the plateau modulus dominates, we predict that the Mooney parameter should drop off rapidly with concentration,

$$C_2(\phi) \sim G_N^* \; \phi^{4/3} \; , \; \nu_x \to 0, \tag{17}$$

while for highly cross-linked rubbers the scaling becomes like that of classical rubber elasticity theory,

$$C_2(\phi) \sim G_c \; \phi^{1/3} \; , \; \nu_x \to \infty \tag{18}$$

The typical effect of network swelling on the elasticity of a natural rubber material swollen by n-decane is nicely discussed by Price and coworkers [62-63] and Mullins and coworkers [8, 64-65] and we refer the reader to this excellent work. Their measurements were performed under a uniaxial extension deformation and deviations from classical elasticity were quantified in terms of the Mooney parameter $C_2$ [See Eq. (15) and (16)], which provides a commonly used measure of deviation from classical elasticity despite the limitation of this parameter to uniaxial extension measurements. This and numerous other experiments have shown that $C_2(\phi)$ is exhibits a concentration power-law with an exponent near 1, although the precise power 4/3 predicted by the LM has been found in careful examinations of the swelling data. [63, 65, 66] The $C_2(\phi)$ term in classical theories of rubber elasticity vanishes so this large change in network elasticity with swelling is simply not accounted for by classical rubber elasticity theories. There are significant implications of this entanglement term for the estimation of solvent interaction parameters of swollen rubbers [67,68] and the cross-link density of rubbery materials [69, 70] which rely on swelling equilibrium measurements and the assumption of classical rubber elasticity theory since these analyses simply neglect a large contribution to the network free energy, an effect noted previously by Mullins. [69] The error involved applying classical theory to the swelling of rubber, such as the Flory-Frenkel-Rehner theory [70,71] should be especially large in highly compliant synthetic rubbery materials and natural rubber materials where the cross-linking level is necessarily low or moderate and chain entanglement interactions prevalent. The very existence of a highly compliant rubbery material normally means that the extent of cross-linking is not hi

In contrast to this lightly cross-linked class of rubbery materials, the localization model indicates that rubbers having a relatively high cross-linking density should exhibit a very different change in their elasticity upon swelling. In these materials, the cross-link contribution in Eq. (11) can dominate the entanglement contribution. In particular, Douglas and McKenna [61] have shown that the weak concentration scaling predicted by Eq. (13), $G_e \sim \phi^{1/3}$, is observed for relatively highly cross-linked natural rubber materials.

Although this concentration dependence is the same as the Frenkel-Flory-Rehner prediction derived from classical rubber elasticity [70,71], the actual functional form of the elasticity is quite different from classical rubber elasticity theory.

## Conclusions

The non-ideal contribution to rubber elasticity can be large in unswollen rubbers, even larger than the contribution arising from chain cross-links. The localization model attributes this non-ideal contribution to strong inter-chain interactions that influence the chain entropy.

- Comparison of localization model to dry rubber deformation data provides a reasonable description of the elasticity of rubbers in all modes of deformation considered so far.

- The Localization model predicts that elasticity of lightly and highly cross-linked rubbers changes in a *qualitatively different* fashion with network dilution.

In view of the simplicity of the analytic form of the model, the physically sensible nature of the parameters derived from it, and its success in capturing qualitative aspects of rubber elasticity in both dry and swollen rubbers, we conclude that the model is a useful working tool in modeling real rubbery materials.

It is argued that complex fluid behavior underlying the entanglement of flexible and stiff polymers, as well as carbon nanotubes, celluse and carbon fibers and even exfoliated clays, involve emergent collective motion governed by physical conditions similar to those which desribe the onset of liquid crystal formation where the complex rheology of these suspensions derives from the entopically-induced self-assembly of particle clumps that form and disintegrate in dynamic equilibrium, a sitution rather similar to the dynamic heterogeneities that arise in glass-forming liquids. We show that the predictions of this entanglement model are broadly in accord with experiments on entangled polymer liquids and this novel viewpoint of entanglement can also naturally explain the relatively enhanced entanglement effects observed in solutions of semi-flexible polymers.

The localization model not only provides an improved physical description of the change in elasticity of rubbery materials with swelling, but it also provides theoretical guidance in the material design of new types of rubber materials. For example, it is well known that if a dry rubbery material is needed exhibiting large and reversible deformation, then the cross-link density cannot be made too large. However, for such compliant materials the dominant, or at least a substantial, contribution to its 'rubbery' consistency derives from chain entanglement; cross-linking is only needed to lock the material into soft equilibrium solid state. Under these conditions, understanding and controlling the molecular structural variables governing entanglement are crucial for the engineering the rubber elasticity. For example, making the chain 'fatter' through the dense grafting of side-groups onto it should make the rubber much softer though its reduction of the plateau modulus $G_N^*$, which in turn dominates the magnitude of entanglement contribution to the rubber elasticity $G_e$ in the low cross-link density limit. Specifically, if the average spatial extension of side-groups of the bottlebrush polymers, which should be on the order of the localization length $\xi$, scales as power-law in the mass $N_s$ of the sides-chains $\xi \sim N_s{}^\nu$, then the entanglement contribution $G_e$ to the dry rubber elasticity of the bottlebrush network should theoretically scale as,

$G_e \sim G_N^* \sim 1 / M_c \sim p^{-3} \sim A^{-3/2}$ [See Eq. (12)] so that $G_e$ should decrease as an inverse power-law, $G_e \sim N_s{}^{-3\nu}$. This means that lightly cross-linked rubbers composed of 'bottlebrush' polymers should be remarkably soft, consistent with the independent suggestions of this trend by Prof. Michael Rubenstein at the Polymer Networks Meeting in Jackson Hole (2012). To achieve this effect, however, it is crucial the polymer backbones are sufficiently extended for these 'fat' polymers to be entangled. This length condition is necessary to achieve the ultra soft rubber rather just a viscous unentangled fluid. A high grafting density of the grafted chains on the polymer chain will also stiffen the chain backbone, and this effect will drive the stiffness of the rubber upward so that there are competing effects in engineering soft rubbery materials through side chain grafting. There are also applications in which the rubber stiffness must not change much solvent exposure and swelling and in which non-linear strain-softening, and associated large deformation instabilities, are not desirable. Simply increasing the cross-link density to high values allows these properties and the sensitivity of the elasticity to the backbone structure of the polymer should be greatly diminished. There seem to be many new opportunities for material design.

## Future Work - Cross-linking History Effects on Rubber Elasticity

There is evidence indicating that networks cross-linked in a highly swollen state exhibit classical rubber elasticity to a good approximation when the rubber is dried. [42]

Since swelling virtually eliminates the entanglement contribution in the localization model, this phenomenology suggests that cross-linking history effects might be modeled by the simple identification $G_N*$(effective) = $G_N(\phi = \phi_x)$ in the localization model and some observations qualitatively support this possibility. [72,73] This would mean that effective entanglement contribution to the network elasticity would equal $G_N*$(effective) = $G_N*$ $(\phi_x / \phi)^2$, where $G_N*$ is the plateau modulus of the equilibrated polymer melt. Subjecting the polymer material to vigorous deformation before cross-linking should also have the effect of delocalizing the chains from the clusters that define the entangled polymer state [12] and this should also lead to a reduction in the entanglement trapping factor $Te$ [5,19] between $G_N*$(effective) and $G_N*$. These possibilities for modifying the entanglement contribution to the network history through changes of the deformation and swelling history of the polymer melt before cross-linking require further investigation.

## Acknowledgement

I would like to thank Richard Gaylord, who pioneered the development of the localization model of rubber elasticity, for many discussions that ultimately led to the current form of the model through our collaboration. The extension of the localization model to describe network swelling owes much to my collaboration with Greg McKenna while he was at NIST and I would not have continued my work in this area without the continued support and discussions on rubber elasticity with Ferenc Horkay.. Finally, I thank Rob Riggleman for providing his illustrative image of the polymer chain configurations before and after deformation obtained from his molecular dynamics simulations of deformed polymer materials.